\documentclass[3p,times,twocolumn]{elsarticle}

\usepackage{ecrc}
\usepackage{braket}

\volume{00}

\firstpage{1}

\journalname{Nuclear and Particle Physics Proceedings}

\runauth{Zhong-Bo Kang, Felix Ringer, Ivan Vitev}

\jid{nppp}

\jnltitlelogo{Nuclear and Particle Physics Proceedings}

\usepackage{amssymb}

\usepackage[figuresright]{rotating}

\begin{document}

\begin{frontmatter}



\dochead{}

\title{Jet and heavy flavor production in heavy-ion collisions}


\author[ucla,zvi,lanl]{Zhong-Bo Kang}
\author[lanl]{Felix Ringer}
\author[lanl]{Ivan Vitev}

\address[ucla]{Department of Physics and Astronomy, University of California, Los Angeles, CA 90095, USA}
\address[zvi]{Mani L. Bhaumik Institute for Theoretical Physics, University of California, Los Angeles, CA 90095, USA}
\address[lanl]{Theoretical Division, Los Alamos National Laboratory, Los Alamos, NM 87545, USA}

\begin{abstract}
We review recent progress in the theoretical description of hard probes in heavy-ion collisions within the framework of Soft Collinear Effective Theory (SCET). Firstly, we consider the inclusive production of heavy flavor mesons and jets with a small radius parameter in proton-proton collisions. Secondly, we describe how in-medium effects can be incorporated consistently with the proton-proton baseline calculations. The in-medium interactions are described by Glauber gluon interactions. We present results for the nuclear modification factor $R_{AA}$ which is most commonly used to study the quenching of hadron or jet production yields in heavy-ion collisions and we compare to recent data from the LHC.
\end{abstract}


\end{frontmatter}


\section{Introduction}

In heavy-ion collisions at the LHC and RHIC, the quark-gluon plasma (QGP) can be reproduced which is predicted to have existed in the early stages of our universe. Highly energetic particles and jets traverse this hot and dense QCD medium and, hence, carry valuable information about its properties. In this work, we focus in particular on open heavy flavor production ($D$- and $B$-mesons) as well as inclusive jet production. In the past years, the experimental collaborations at the LHC and RHIC have provided high precision data for the nuclear modification factor $R_{AA}$, see for example~\cite{Adare:2010de,Adamczyk:2014uip,CMS:2016nrh,Khachatryan:2016ypw,Aad:2012vca,Abelev:2013fn,Khachatryan:2016jfl}.

In order to study the modification of open heavy flavor yields in heavy-ion collisions, we rely on recently developed techniques within Soft Collinear Effective Theory (SCET)~\cite{Bauer:2000yr,Bauer:2001yt}. For brief overview of recent SCET developments in hadronic and nuclear collisions see~\cite{ivitev}. In order to describe the in-medium interactions of energetic partons inside the QGP, additional Glauber modes are included at the level of the SCET Lagrangian~\cite{Idilbi:2008vm,Ovanesyan:2011xy}. The corresponding effective theory for gluons and massive quarks propagating in the QGP is referred to as SCET$_{\mathrm{M,G}}$~\cite{Kang:2016ofv}. Within this framework, it is possible to calculate the full collinear in-medium splitting functions which allows us to go beyond the traditional approach to parton energy loss~\cite{Gyulassy:2000er,Djordjevic:2003zk}. In this work, we present a novel implementation using these in-medium collinear splitting functions which is consistent with next-to-leading order (NLO) calculations used for the proton-proton ($pp$) baseline.

The second topic addressed in this work is the description of inclusive jet spectra. In the era of the LHC and RHIC, the measurement of jets in heavy-ion collisions and their theoretical description have received an increasing attention both from the theoretical~\cite{He:2011pd,Chien:2014nsa,Huang:2015mva,Wang:2016fds} and the experimental communities~\cite{Aad:2012vca,Abelev:2013fn,Khachatryan:2016jfl}. The cross section for jets crucially depends on the jet size parameter $R$. Typically, $R$ is chosen in the range of $R\sim 0.2-0.4$ by the experimental analyses in order to avoid fluctuations in the heavy-ion background. The perturbative calculations exhibit a single logarithmic structure $\alpha_s^n\ln^n R$ which needs to be resummed to all orders~\cite{Kang:2016mcy,Kang:2016ehg}. The results presented here for the $pp$ baseline will eventually be the starting point for describing small-$R$ jets in the medium.

\section{Heavy flavor production}

\begin {figure*}[t]
\begin{center}
\vspace*{10mm}
\includegraphics[width=0.4\textwidth,trim=1cm 2cm 1cm 1cm ]{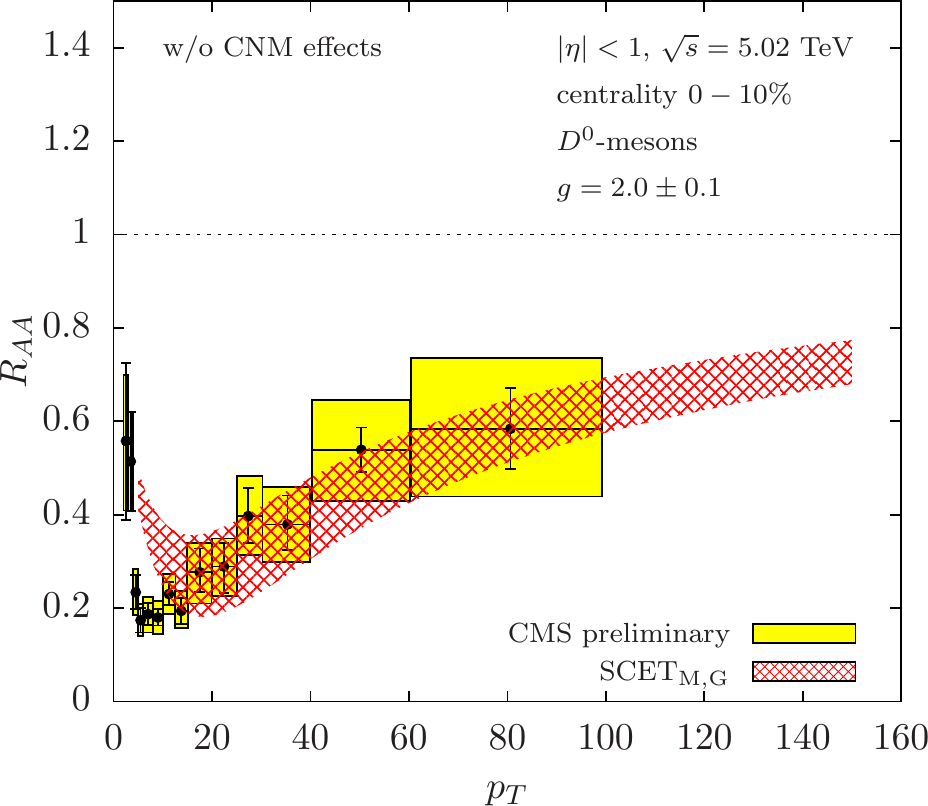} 
\hspace*{2cm}
\includegraphics[width=0.4\textwidth,trim=1cm 2cm 1cm 1cm ]{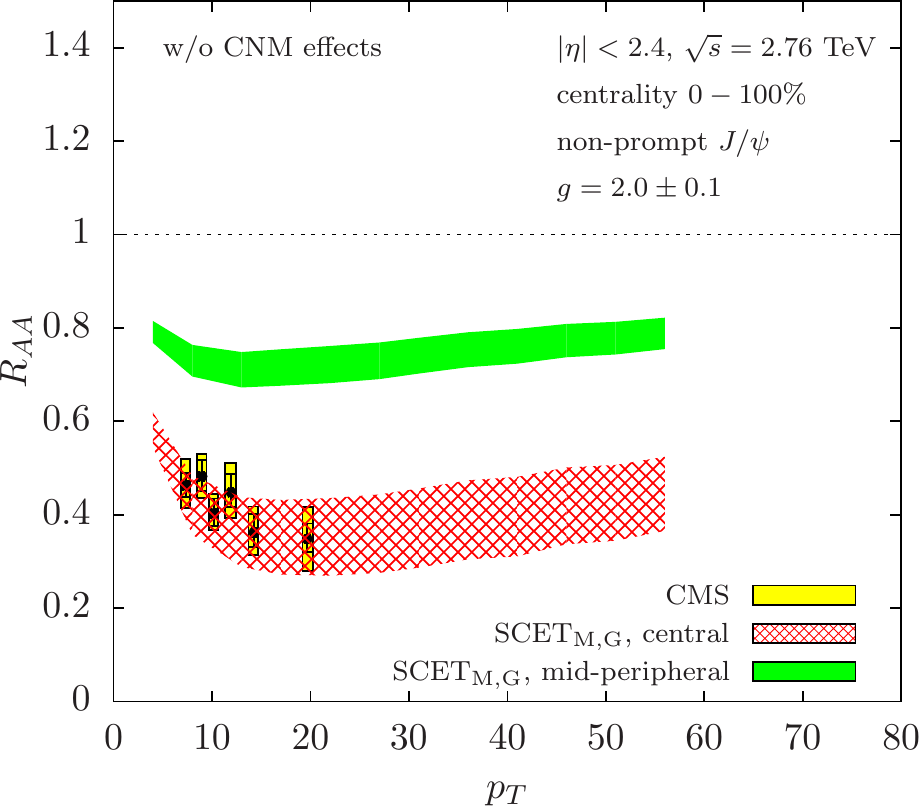} 
\end{center}
\vspace*{1.4cm}
\caption{\label{fig:HF} Suppression of $D^0$ mesons~\cite{CMS:2016nrh} (left) and non-prompt $J/\psi$ which originate from $B$-hadrons~\cite{Khachatryan:2016ypw} (right) in $\mathrm{PbPb}$ collisions at the LHC. The data was taken by the CMS collaboration. Our results obtained within SCET$_{\mathrm{M,G}}$ are shown for central collisions (red hatched band) and for mid-peripheral collisions (green band). Here, we do not include Cold Nuclear Matter (CNM) effects.}
\end{figure*}

We calculate the $pp$ baseline for open heavy flavor production using the Zero Mass Variable Flavor Number Scheme (ZMVFNS) at NLO and the fragmentation functions (FFs) of~\cite{Kneesch:2007ey,Kniehl:2008zza}. We find a good description of the $pp$ data even for relatively small transverse momenta $p_T\sim 5$~GeV. The part of the NLO cross section that describes the fragmentation of a parton $i$ into a hadron $H$ can schematically be written as
\begin{equation}
\sum_j\sigma_i^{(0)}\otimes{\cal P}_{i\to jk}\otimes D_j^H \, .
\end{equation}
Here $\sigma_i^{(0)}$ is the leading-order production cross section for a parton $i$, ${\cal P}_{i\to jk}$ describes the splitting process $i\to jk$ and $D_j^H$ is a non-perturbative fragmentation function. In the medium, we have to take into account both vacuum and medium induced parton emissions. The description of the splitting process gets modified as
\begin{equation}
{\cal P}_{i\to jk} = {\cal P}^{\mathrm{vac}}_{i\to jk} + {\cal P}^{\mathrm{med}}_{i\to jk} \, ,
\end{equation}
where ${\cal P}^{\mathrm{med}}_{i\to jk}$ is the collinear in-medium splitting function which can be obtained from SCET$_{\mathrm{M,G}}$. Using power counting techniques, we find that the required effective field theory including both finite quark masses and Glauber gluons describing the interaction with the medium is given by a direct sum of the massive vacuum SCET Lagrangian~\cite{Leibovich:2003jd} and the massless in-medium interaction terms~\cite{Ovanesyan:2011xy}. See~\cite{Kang:2016ofv} for a more detailed discussion. To first order in opacity~\cite{Gyulassy:2000er}, we can compute the diagonal splitting process $Q\to Qg$ as well as the off-diagonal ones $Q\to gQ$ and $g\to Q\bar Q$ in the medium. In the soft emission limit, we recover the results obtained within the traditional picture of parton energy loss for massive quarks~\cite{Djordjevic:2003zk}.

\begin {figure*}[t]
\begin{center}
\vspace*{10mm}
\includegraphics[width=0.8\textwidth,trim=1cm 2cm 1cm 1cm ]{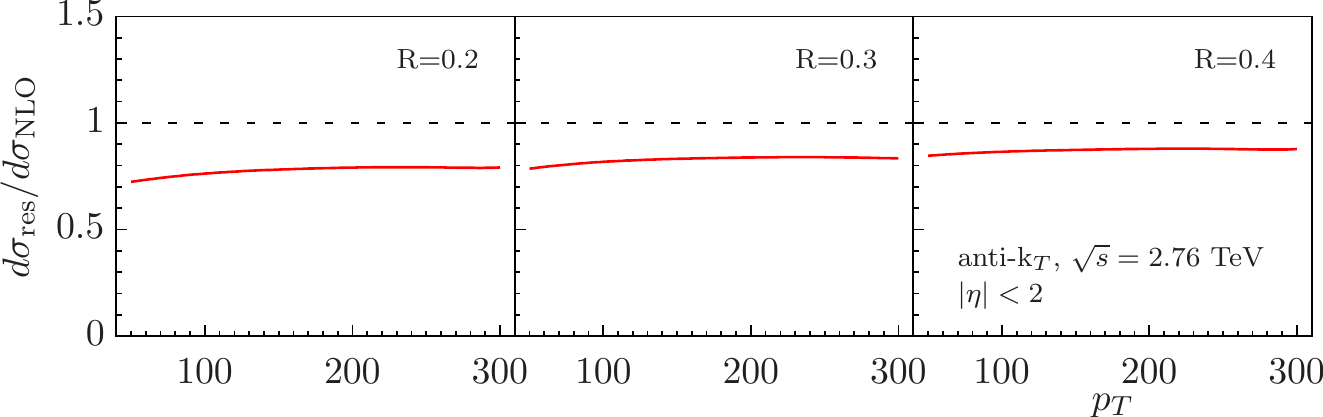} 
\end{center}
\vspace*{1.6cm}
\caption{\label{fig:jets} Ratio of the $\ln R$ resummed jet cross section and the NLO result in proton-proton collisions for $\sqrt{s}=2.76$~TeV, $|\eta|<2$ using the anti-$k_T$ reclustering algorithm for $R=0.2,\, 0.3,\, 0.4$ as in the CMS analysis of~\cite{Khachatryan:2016jfl}.}
\end{figure*}
\vspace*{-.2cm}

In Fig.~\ref{fig:HF}, we present numerical results for the nuclear modification factor
\begin{equation}
R_{AA}=\frac{d\sigma_{\mathrm{PbPb}}^H/d\eta dp_T }{\braket{N_{\mathrm{coll}}} d\sigma^H_{pp}/d\eta dp_T} \, ,
\end{equation}
where $\braket{N_{\mathrm{coll}}}$ is the average number of binary nucleon-nucleon collisions. On the left side, we present results using the new framework for $D^0$ mesons (hatched red band) and we find good agreement compared to the CMS data of~\cite{CMS:2016nrh} taken at $\sqrt{s}=5.02$~TeV, $|\eta|<1$ and $0-10\%$ centrality. Similarly, on the right side of Fig.~\ref{fig:HF}, we show results for the suppression of non-prompt $J/\psi$ which originate from $B$-hadrons and we compare to the CMS data of~\cite{Khachatryan:2016ypw} ($\sqrt{s}=2.76$~TeV, $|\eta|<2.4$). Note that the non-prompt $J/\psi$ data is not for fixed centrality but for minimum bias events. Therefore, we present our calculation for both central (hatched red band) and mid-peripheral collisions (green band). The experimental minimum bias results are dominated by central collisions and we find good agreement with our calculation for $0-10\%$ centrality. We would like to point out that the suppression rates crucially depend on whether the observed heavy meson is produced by a heavy quark or a gluon since gluons lose more energy in the medium.

\section{Small jet radius resummation}

Within the framework of SCET, we define the so-called ``semi-inclusive jet functions'' (siJFs) $J_i(z,R,\mu)$ that describe the transition of a final state parton $i$ into a jet with radius $R$ carrying an energy fraction $z$ of the parent parton~\cite{Kang:2016mcy}. It turns out that there is a close analogy between the siJFs and standard collinear FFs. When considering the inclusive production cross section of jets or hadrons, the same factorized framework can be used where the siJF/ FFs appear convolved with the same hard-scattering functions. For the double differential cross section for inclusive jets in the rapidity $\eta$ and the transverse momentum $p_T$ of the jet, we find schematically
\begin{equation}
\frac{d\sigma^{\mathrm{jet}}}{d\eta dp_T}=\sum_{abc} f_a\otimes f_b\otimes H_{ab}^c\otimes J_c \, .
\end{equation}
Here $f_{a,b}$ denote the PDFs, $H_{ab}^c$ are the hard-scattering functions and $J_c$ are the siJFs. See also~\cite{Jager:2004jh,Kaufmann:2015hma}. In addition, the siJFs satisfy the same timelike DGLAP evolution equations as FFs. We have
\begin{eqnarray}
&&\mu\frac{d}{d\mu}J_i(z,R,\mu) = \nonumber \\
&&\hspace*{-1cm} \frac{\alpha_s(\mu)}{\pi}\sum_j\int_z^1\frac{dz'}{z'}P_{ji}\left(\frac{z}{z'}\right)J_j(z',R,\mu) \, ,
\end{eqnarray}
where the anomalous dimensions $P_{ji}(z)$ are the usual Altarelli-Parisi splitting kernels. When solving the DGLAP equations for the siJFs, single logarithms in the jet radius parameter $R$ can be resummed. Note that the jet functions in these evolution equations contain distributions in $1-z$. Therefore, we solve the DGLAP equations in Mellin moment space using the numerical methods developed in~\cite{Vogt:2004ns,Anderle:2015lqa}. We are able to achieve a precision of next-to-leading order combined with next-to-leading-logarithmic accuracy (NLO+NLL$_R$). The resummation of terms $\sim\ln R$ turns out to be significant for precision calculations even for relatively large values of $R\sim 0.6$. Jet cross sections based on fixed order calculations can have a vanishing, unphysical QCD scale dependence in some kinematic regions~\cite{Dasgupta:2014yra}. Instead, when including the resummation of $\ln R$ terms, a finite size of the QCD scale uncertainty band is obtained allowing a realistic estimate of unknown higher order corrections.

Recently, the CMS collaboration reported precise measurements of inclusive jet spectra for several different jet radii $R=0.2,\, 0.3, \,0.4$~\cite{Khachatryan:2016jfl}. The data was taken at $\sqrt{s}=2.76$~TeV and $|\eta|<2$ using the anti-$k_T$ reclustering algorithm~\cite{Cacciari:2008gp}. It turns out that the data is not well described by NLO calculations. Typically, PDF fits rely on inclusive jet data but at larger $R$. In Fig.~\ref{fig:jets}, we plot for the same kinematics the ratio of the $\ln R$ resummed cross section and the NLO result. This ratio matches the data/ NLO ratio in~\cite{Khachatryan:2016jfl} and confirms our predictions for small-$R$ jets in~\cite{Kang:2016mcy}. Therefore, the $\ln R$ resummed calculations constitute the ideal starting point in order to consistently derive the modification of jet production yields in heavy-ion collisions.

\section{Conclusions}

We have presented results within SCET$_{\mathrm{M,G}}$ for the medium modification of open heavy flavor production in heavy-ion collisions. We found good agreement with data taken at the LHC for both $D$- and $B$-mesons. In addition, we presented a new framework for inclusive small-$R$ jets in $pp$ collisions allowing for the resummation of single logarithms in the jet size parameter $R$ at NLL$_R$ accuracy. We found that our predictions match recent data from the LHC. In the future, we are going to extend our calculations to inclusive jets in heavy-ion collisions. In addition, it will be possible to address the modification of jet substructure observables~\cite{Kang:2016ehg}.

\section*{Acknowledgements}

This work is supported by the U.S. Department of Energy under Contract No. DE-AC52-06NA25396, in part by the LDRD program at Los Alamos National Laboratory.




\nocite{*}
\bibliographystyle{elsarticle-num}
\bibliography{Ringer_F}

\begin{thebibliography}{10}
\expandafter\ifx\csname url\endcsname\relax
  \def\url#1{\texttt{#1}}\fi
\expandafter\ifx\csname urlprefix\endcsname\relax\def\urlprefix{URL }\fi
\expandafter\ifx\csname href\endcsname\relax
  \def\href#1#2{#2} \def\path#1{#1}\fi

\bibitem{Adare:2010de}
A.~Adare, et~al., {Heavy Quark Production in $p+p$ and Energy Loss and Flow of
  Heavy Quarks in Au+Au Collisions at $\sqrt{s_{NN}}=200$ GeV}, Phys. Rev. C84
  (2011) 044905.
\newblock \href {http://arxiv.org/abs/1005.1627} {\path{arXiv:1005.1627}},
  \href {http://dx.doi.org/10.1103/PhysRevC.84.044905}
  {\path{doi:10.1103/PhysRevC.84.044905}}.

\bibitem{Adamczyk:2014uip}
L.~Adamczyk, et~al., {Observation of $D^0$ Meson Nuclear Modifications in Au+Au
  Collisions at $\sqrt{s_{NN}}=200$  GeV}, Phys. Rev. Lett. 113~(14) (2014)
  142301.
\newblock \href {http://arxiv.org/abs/1404.6185} {\path{arXiv:1404.6185}},
  \href {http://dx.doi.org/10.1103/PhysRevLett.113.142301}
  {\path{doi:10.1103/PhysRevLett.113.142301}}.

\bibitem{CMS:2016nrh}
C.~Collaboration, {$\mathrm{D}^0$ meson nuclear modification factor in PbPb
  collisions at $\sqrt{s_\mathrm{NN}} = 5.02~\mathrm{TeV}$}.

\bibitem{Khachatryan:2016ypw}
V.~Khachatryan, et~al., {Suppression and azimuthal anisotropy of prompt and
  nonprompt $J/\psi$ production in PbPb collisions at $\sqrt{s_{NN}}$ = 2.76
  TeV}, Eur. Phys. J. C\href {http://arxiv.org/abs/1610.00613}
  {\path{arXiv:1610.00613}}, \href {http://dx.doi.org/10.3204/PUBDB-2016-04916}
  {\path{doi:10.3204/PUBDB-2016-04916}}.

\bibitem{Aad:2012vca}
G.~Aad, et~al., {Measurement of the jet radius and transverse momentum
  dependence of inclusive jet suppression in lead-lead collisions at
  $\sqrt{s_{NN}}$= 2.76 TeV with the ATLAS detector}, Phys. Lett. B719 (2013)
  220--241.
\newblock \href {http://arxiv.org/abs/1208.1967} {\path{arXiv:1208.1967}},
  \href {http://dx.doi.org/10.1016/j.physletb.2013.01.024}
  {\path{doi:10.1016/j.physletb.2013.01.024}}.

\bibitem{Abelev:2013fn}
B.~Abelev, et~al., {Measurement of the inclusive differential jet cross section
  in $pp$ collisions at $\sqrt{s} = 2.76$ TeV}, Phys. Lett. B722 (2013)
  262--272.
\newblock \href {http://arxiv.org/abs/1301.3475} {\path{arXiv:1301.3475}},
  \href {http://dx.doi.org/10.1016/j.physletb.2013.04.026}
  {\path{doi:10.1016/j.physletb.2013.04.026}}.

\bibitem{Khachatryan:2016jfl}
V.~Khachatryan, et~al., {Measurement of inclusive jet cross-sections in pp and
  PbPb collisions at $\sqrt{s_{NN}}$=2.76 TeV}, Submitted to: Phys. Rev. C\href
  {http://arxiv.org/abs/1609.05383} {\path{arXiv:1609.05383}}.

\bibitem{Bauer:2000yr}
C.~W. Bauer, S.~Fleming, D.~Pirjol, I.~W. Stewart, {An Effective field theory
  for collinear and soft gluons: Heavy to light decays}, Phys. Rev. D63 (2001)
  114020.
\newblock \href {http://arxiv.org/abs/hep-ph/0011336}
  {\path{arXiv:hep-ph/0011336}}, \href
  {http://dx.doi.org/10.1103/PhysRevD.63.114020}
  {\path{doi:10.1103/PhysRevD.63.114020}}.

\bibitem{Bauer:2001yt}
C.~W. Bauer, D.~Pirjol, I.~W. Stewart, {Soft collinear factorization in
  effective field theory}, Phys. Rev. D65 (2002) 054022.
\newblock \href {http://arxiv.org/abs/hep-ph/0109045}
  {\path{arXiv:hep-ph/0109045}}, \href
  {http://dx.doi.org/10.1103/PhysRevD.65.054022}
  {\path{doi:10.1103/PhysRevD.65.054022}}.

\bibitem{ivitev}
I.~Vitev, these proceedings.

\bibitem{Idilbi:2008vm}
A.~Idilbi, A.~Majumder, {Extending Soft-Collinear-Effective-Theory to describe
  hard jets in dense QCD media}, Phys. Rev. D80 (2009) 054022.
\newblock \href {http://arxiv.org/abs/0808.1087} {\path{arXiv:0808.1087}},
  \href {http://dx.doi.org/10.1103/PhysRevD.80.054022}
  {\path{doi:10.1103/PhysRevD.80.054022}}.

\bibitem{Ovanesyan:2011xy}
G.~Ovanesyan, I.~Vitev, {An effective theory for jet propagation in dense QCD
  matter: jet broadening and medium-induced bremsstrahlung}, JHEP 06 (2011)
  080.
\newblock \href {http://arxiv.org/abs/1103.1074} {\path{arXiv:1103.1074}},
  \href {http://dx.doi.org/10.1007/JHEP06(2011)080}
  {\path{doi:10.1007/JHEP06(2011)080}}.

\bibitem{Kang:2016ofv}
Z.-B. Kang, F.~Ringer, I.~Vitev, {Effective field theory approach to open heavy
  flavor production in heavy-ion collisions}\href
  {http://arxiv.org/abs/1610.02043} {\path{arXiv:1610.02043}}.

\bibitem{Gyulassy:2000er}
M.~Gyulassy, P.~Levai, I.~Vitev, {Reaction operator approach to nonAbelian
  energy loss}, Nucl. Phys. B594 (2001) 371--419.
\newblock \href {http://arxiv.org/abs/nucl-th/0006010}
  {\path{arXiv:nucl-th/0006010}}, \href
  {http://dx.doi.org/10.1016/S0550-3213(00)00652-0}
  {\path{doi:10.1016/S0550-3213(00)00652-0}}.

\bibitem{Djordjevic:2003zk}
M.~Djordjevic, M.~Gyulassy, {Heavy quark radiative energy loss in QCD matter},
  Nucl. Phys. A733 (2004) 265--298.
\newblock \href {http://arxiv.org/abs/nucl-th/0310076}
  {\path{arXiv:nucl-th/0310076}}, \href
  {http://dx.doi.org/10.1016/j.nuclphysa.2003.12.020}
  {\path{doi:10.1016/j.nuclphysa.2003.12.020}}.

\bibitem{He:2011pd}
Y.~He, I.~Vitev, B.-W. Zhang, {${\cal O}(\alpha_s^3)$ Analysis of Inclusive Jet
  and di-Jet Production in Heavy Ion Reactions at the Large Hadron Collider},
  Phys. Lett. B713 (2012) 224--232.
\newblock \href {http://arxiv.org/abs/1105.2566} {\path{arXiv:1105.2566}},
  \href {http://dx.doi.org/10.1016/j.physletb.2012.05.054}
  {\path{doi:10.1016/j.physletb.2012.05.054}}.

\bibitem{Chien:2014nsa}
Y.-T. Chien, I.~Vitev, {Jet Shape Resummation Using Soft-Collinear Effective
  Theory}, JHEP 12 (2014) 061.
\newblock \href {http://arxiv.org/abs/1405.4293} {\path{arXiv:1405.4293}},
  \href {http://dx.doi.org/10.1007/JHEP12(2014)061}
  {\path{doi:10.1007/JHEP12(2014)061}}.

\bibitem{Huang:2015mva}
J.~Huang, Z.-B. Kang, I.~Vitev, H.~Xing, {Photon-tagged and B-meson-tagged
  b-jet production at the LHC}, Phys. Lett. B750 (2015) 287--293.
\newblock \href {http://arxiv.org/abs/1505.03517} {\path{arXiv:1505.03517}},
  \href {http://dx.doi.org/10.1016/j.physletb.2015.09.029}
  {\path{doi:10.1016/j.physletb.2015.09.029}}.

\bibitem{Wang:2016fds}
X.-N. Wang, S.-Y. Wei, H.-Z. Zhang, {Effect of medium recoil and $p_T$
  broadening on single inclusive jet suppression in high-energy heavy-ion
  collisions}\href {http://arxiv.org/abs/1611.07211} {\path{arXiv:1611.07211}}.

\bibitem{Kang:2016mcy}
Z.-B. Kang, F.~Ringer, I.~Vitev, {The semi-inclusive jet function in SCET and
  small radius resummation for inclusive jet production}, JHEP 10 (2016) 125.
\newblock \href {http://arxiv.org/abs/1606.06732} {\path{arXiv:1606.06732}},
  \href {http://dx.doi.org/10.1007/JHEP10(2016)125}
  {\path{doi:10.1007/JHEP10(2016)125}}.

\bibitem{Kang:2016ehg}
Z.-B. Kang, F.~Ringer, I.~Vitev, {Jet substructure using semi-inclusive jet
  functions in SCET}, JHEP 11 (2016) 155.
\newblock \href {http://arxiv.org/abs/1606.07063} {\path{arXiv:1606.07063}},
  \href {http://dx.doi.org/10.1007/JHEP11(2016)155}
  {\path{doi:10.1007/JHEP11(2016)155}}.

\bibitem{Kneesch:2007ey}
T.~Kneesch, B.~A. Kniehl, G.~Kramer, I.~Schienbein, {Charmed-meson
  fragmentation functions with finite-mass corrections}, Nucl. Phys. B799
  (2008) 34--59.
\newblock \href {http://arxiv.org/abs/0712.0481} {\path{arXiv:0712.0481}},
  \href {http://dx.doi.org/10.1016/j.nuclphysb.2008.02.015}
  {\path{doi:10.1016/j.nuclphysb.2008.02.015}}.

\bibitem{Kniehl:2008zza}
B.~A. Kniehl, G.~Kramer, I.~Schienbein, H.~Spiesberger, {Finite-mass effects on
  inclusive $B$ meson hadroproduction}, Phys. Rev. D77 (2008) 014011.
\newblock \href {http://arxiv.org/abs/0705.4392} {\path{arXiv:0705.4392}},
  \href {http://dx.doi.org/10.1103/PhysRevD.77.014011}
  {\path{doi:10.1103/PhysRevD.77.014011}}.

\bibitem{Leibovich:2003jd}
A.~K. Leibovich, Z.~Ligeti, M.~B. Wise, {Comment on quark masses in SCET},
  Phys. Lett. B564 (2003) 231--234.
\newblock \href {http://arxiv.org/abs/hep-ph/0303099}
  {\path{arXiv:hep-ph/0303099}}, \href
  {http://dx.doi.org/10.1016/S0370-2693(03)00565-3}
  {\path{doi:10.1016/S0370-2693(03)00565-3}}.

\bibitem{Jager:2004jh}
B.~Jager, M.~Stratmann, W.~Vogelsang, {Single inclusive jet production in
  polarized $p p$ collisions at ${\cal O}(\alpha^3_s)$}, Phys. Rev. D70 (2004)
  034010.
\newblock \href {http://arxiv.org/abs/hep-ph/0404057}
  {\path{arXiv:hep-ph/0404057}}, \href
  {http://dx.doi.org/10.1103/PhysRevD.70.034010}
  {\path{doi:10.1103/PhysRevD.70.034010}}.

\bibitem{Kaufmann:2015hma}
T.~Kaufmann, A.~Mukherjee, W.~Vogelsang, {Hadron Fragmentation Inside Jets in
  Hadronic Collisions}, Phys. Rev. D92~(5) (2015) 054015.
\newblock \href {http://arxiv.org/abs/1506.01415} {\path{arXiv:1506.01415}},
  \href {http://dx.doi.org/10.1103/PhysRevD.92.054015}
  {\path{doi:10.1103/PhysRevD.92.054015}}.

\bibitem{Vogt:2004ns}
A.~Vogt, {Efficient evolution of unpolarized and polarized parton distributions
  with QCD-PEGASUS}, Comput. Phys. Commun. 170 (2005) 65--92.
\newblock \href {http://arxiv.org/abs/hep-ph/0408244}
  {\path{arXiv:hep-ph/0408244}}, \href
  {http://dx.doi.org/10.1016/j.cpc.2005.03.103}
  {\path{doi:10.1016/j.cpc.2005.03.103}}.

\bibitem{Anderle:2015lqa}
D.~P. Anderle, F.~Ringer, M.~Stratmann, {Fragmentation Functions at
  Next-to-Next-to-Leading Order Accuracy}, Phys. Rev. D92~(11) (2015) 114017.
\newblock \href {http://arxiv.org/abs/1510.05845} {\path{arXiv:1510.05845}},
  \href {http://dx.doi.org/10.1103/PhysRevD.92.114017}
  {\path{doi:10.1103/PhysRevD.92.114017}}.

\bibitem{Dasgupta:2014yra}
M.~Dasgupta, F.~Dreyer, G.~P. Salam, G.~Soyez, {Small-radius jets to all orders
  in QCD}, JHEP 04 (2015) 039.
\newblock \href {http://arxiv.org/abs/1411.5182} {\path{arXiv:1411.5182}},
  \href {http://dx.doi.org/10.1007/JHEP04(2015)039}
  {\path{doi:10.1007/JHEP04(2015)039}}.

\bibitem{Cacciari:2008gp}
M.~Cacciari, G.~P. Salam, G.~Soyez, {The Anti-k(t) jet clustering algorithm},
  JHEP 04 (2008) 063.
\newblock \href {http://arxiv.org/abs/0802.1189} {\path{arXiv:0802.1189}},
  \href {http://dx.doi.org/10.1088/1126-6708/2008/04/063}
  {\path{doi:10.1088/1126-6708/2008/04/063}}.

\end{thebibliography}







\end{document}